# STUDY OF EFFECTS OF FAILURE OF BEAMLINE ELEMENTS & THEIR COMPENSATION IN CW SUPERCONDUCTING LINAC *

A. Saini[#], K. Ranjan, University of Delhi, Delhi, India
N. Solyak, S. Mishra, V. Yakovlev, FNAL, Batavia, IL 60510, U.S.A.

*Abstract*

Project-X is the proposed high intensity proton facility to be built at Fermilab, US. First stage of the Project-X consists of superconducting linac which will be operated in continuous wave (CW) mode to accelerate the beam from 2.5 MeV to 3 GeV. The operation at CW mode puts high tolerances on the beam line components, particularly on radiofrequency (RF) cavity. The failure of beam line elements at low energy is very critical as it results in mis-match of the beam with the following sections due to different beam parameters than designed parameter. It makes the beam unstable which causes emittance dilution, and ultimately results in beam losses. In worst case, it could affect the reliability of the machine and may lead to the shutdown of the Linac to replace the failed elements. Thus, it is important to study these effects and their compensation to get smooth beam propagation in linac. This paper describes the results of study performed for the failure of RF cavity & solenoid in SSR0 section.

## INTRODUCTION

Project-X is a high intensity multi megawatt (MW) facility (Fig .1) to be built at Fermilab [1].

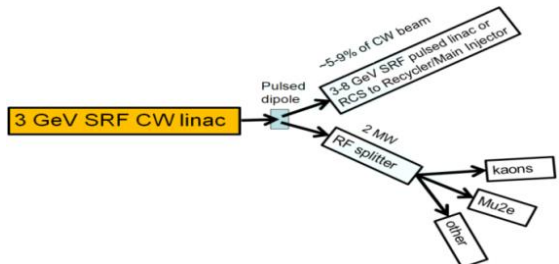

Figure. 1: Project-X facility configuration

The proposed facility is based on 3 GeV, 1mA CW SC linac [2]. The schematic of baseline configuration of the linac is shown in Fig. 2. It includes an ion source which provides 5 mA pulsed beam of H[-] ions. The beam is accelerated through the RFQ which is operated at room temperature at 325 MHz frequency. The RFQ is followed by Medium Energy Beam Transport (MEBT) section which is used to chop the beam in order to get the time structure which is necessary to operate the different experiments simultaneously. The MEBT is followed by SC linac, which is segmented into two sections: low energy part and high energy part. The low energy section (2.5-160 MeV) uses three families of SC single spoke resonators i.e. SSR0, SSR1 & SSR2 which are operated at 325 MHz. The high energy section of the SC linac (160 MeV-3.0 GeV) uses two families of 5 cell SC elliptical shape cavities i.e. $\beta=0.61$ and $\beta=0.9$ which are operated at 650 MHz. Numbers of beam line elements (RF cavities, solenoids and quadrupoles) in each section along with their transition points are summarized in Table 1.

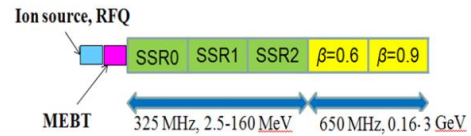

Figure. 2: Acceleration scheme for CW linac.

Table 1: No. of elements along with transition energy.

|  | SSR0 | SSR1 | SSR2 | $\beta=0.6$ | $\beta=0.9$ |
|---|---|---|---|---|---|
| Cavities | 18 | 20 | 40 | 36 | 152 |
| Solenoids | 18 | 20 | 20 | 0 | 0 |
| Quads | 0 | 0 | 0 | 24 | 38 |
| CM | 1 | 2 | 4 | 6 | 19 |
| Period length (m) | 0.61 | 0.80 | 1.60 | 5.00 | 15.40 |
| Section length (m) | 10.98 | 16.40 | 33.20 | 60.00 | 292.60 |
| Transition energy, MeV | 10.18 | 42.58 | 160.5 | 515.4 | 3028.3 |

Operation of the SC linac at CW mode puts high tolerances on beam transport elements, particularly on RF cavities. Failure of the beam transport elements like cavity, solenoid and quadrupole alters the focusing period of the beam, resulting in a mismatch of the beam with the subsequent sections. This, in turn, causes beam losses. In some cases, failure of the beam transport element results in complete beam losses and it becomes necessary to replace this element for nominal operation of the machine. These beam interruptions reduce the beam availability for the different experiments. To achieve high performance and hence reliability of the machine, it is necessary to include this scenario during the design of the SC linac lattice. It should have a capability that RF cavity or magnet failure could be compensated locally by using the neighbouring elements. In this paper, we demonstrate the compensation of failure of RF cavity and solenoid for the most critical case of failure of first cavity and first solenoid in SSR0 section. We would also like to emphasize that even though the latest design of SC CW linac [2] differs marginally, however studies presented in this paper can be taken as guideline.

## FAILURE OF RF CAVITY

Failure of a cavity components (tuner, power coupler, etc.), degradation of RF cavity surface during operation are such scenarios which could be results in degradation of performance of RF cavity. Calculations have been performed to study the effects of failure of RF cavity, especially in the low energy section of the SC linac where beam is non-relativistic and space charge effect dominates. Due to cavity failure, significant deformation in the longitudinal phase space takes place, which in turn induces strong envelope oscillations and halo formation. Through the coupling, a significant enhancement of halo formation in the transverse phase space is also introduced.

___________________
*Work supported by IUSSTF and US DOE
[#]asaini@fnal.gov

*Failure of first RF cavity in SSR0 section*

Analysis is done to compensate the effects of failure of first RF cavity in SSR0 section with minimal user disruption. Neighbouring elements are retuned in order to achieve designed beam energy and smooth beam propagation through the linac. It is considered most critical case as beam energy is very low and only few elements are available to retune the lattice. It is ensured that maximum surface magnetic field in SSR0 cavity should be below 60 mT which is a limitation for spoke resonators in 325 MHz section. Another constraint is the magnetic field in solenoid that should be less than 6 T/m.

Simulation is performed using multi-particle tracking code TRACEWIN (PARTRAN) [3]. Beam r.m.s. envelopes and normalized emittances are shown in Fig. 3 calculated from tracking of 10k macro-particles through nominal SC linac lattice (no misalignment errors and no failure of beam line elements).

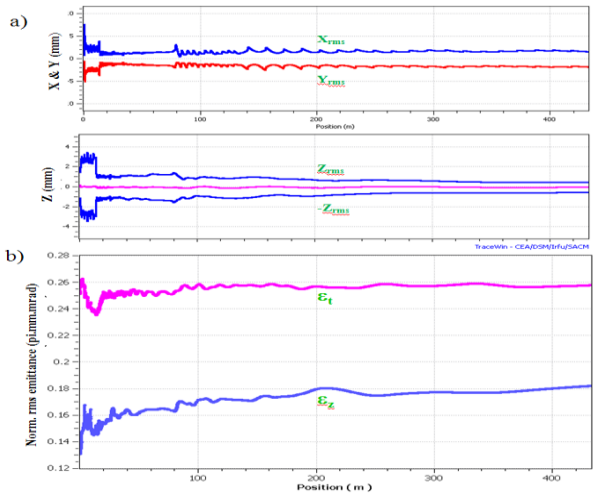

Figure 3: Transverse (x, y) and Longitudinal (a) r.m.s. beam envelope ($x_{rms}$, $y_{rms}$ & $z_{rms}$) and (b) normalized Emittance after tracking through the SC linac without any failure of beam transport elements.

Figure 4 shows the beam profile and normalized emittance through the SC linac in case of the failure of first cavity in SSR0 section. It can be seen that longitudinal profile of the beam blows up which results in emittance dilution and increase in beam losses. Significant beam losses mostly happened in transition between sections causes reduction of longitudinal emittance and beam size. Transverse profile of the beam remains unaffected from failure effects of rf cavity. Thus, in this situation to operate the SC linac without beam losses it is necessary to compensate this effect.

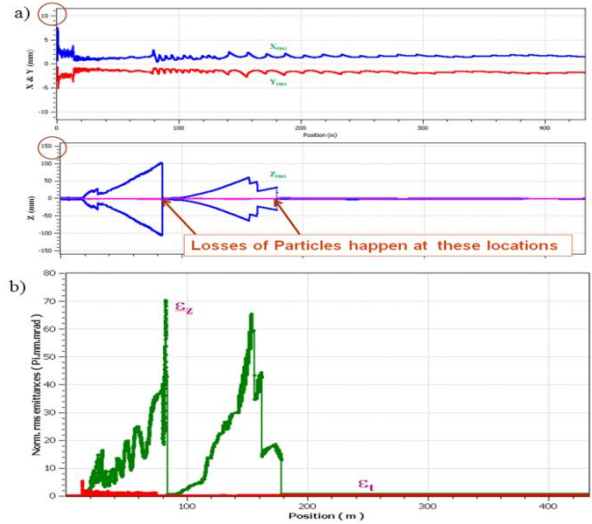

Figure 4: Transverse (x, y) and Longitudinal (a) r.m.s. beam envelope and (b) Emittance after tracking through the SC linac with failure of first cavity in SSR0 section.

In the compensation scheme a few elements in the vicinity of failed elements are re-adjusted (RF phase and field amplitude of RF cavity and field gradient of the solenoid and quad) to achieve designed energy and smooth beam propagation through the SC linac. All the cavities in the SC linac are operated with independent RF sources, which provide the freedom of varying the RF phase and field amplitude of each cavity separately. Fig. 5 shows the beam profile (zoomed region around the failed cavity) after compensation of the failed cavity. Neighbouring beam transport elements used for re-adjustment are highlighted in green colour: two cavities (referred as Gaps), two solenoids and a quad triplet in the upstream MEBT section, one solenoid in the same period of the failed cavity, and one solenoid and one cavity in each of the two downstream periods (after the failed cavity). The final emiitance dilution after compensation is shown in Figure 6. It can be easily seen that emittance is improved significantly and no beam losses happen after applying compensation scheme.

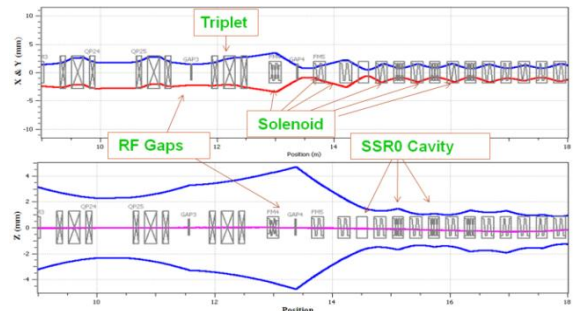

Figure 5: Beam profile in transverse plane (top) & longitudinal plane (bottom) in presence of a failed cavity after its compensation.

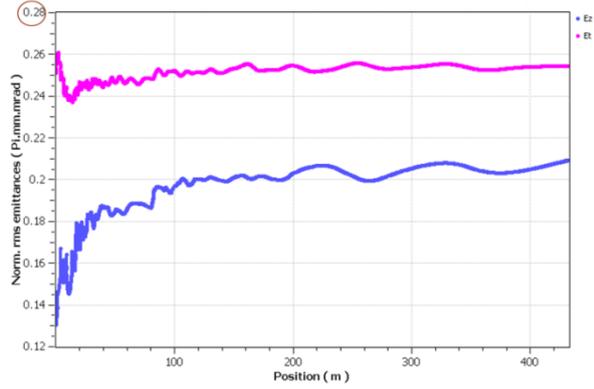

Figure. 6: Longitudinal (blue) and transverse (magenta) emittance dilution through SC linac in presence of failed cavity with compensation scheme.

## FAILURE OF TRANSVERSE FOCUSING ELEMENTS

Study of failure of transverse focusing elements (Solenoids & Quads) is also performed. It is found that failure of first focusing element (solenoid) in SC CW linac is as critical as failure of first RF cavity in SSR0 section. It results mismatch in transverse plane and leads emittance dilution and hence beam losses. Fig. 7 beam profile and normalized emittances along the linac after failure of first solenoid in SSR0 section. Compensation scheme is applied to mitigate failure's effect and achieving smooth beam propagation through the linac. Two cavities (referred as Gaps), two solenoids and a quad triplet in the upstream MEBT section, one cavity in the same period of the failed solenoid, and one solenoid and one cavity in each of the two downstream periods (after the failed solenoid) are used to retune the beam optics. It can be observed from Fig. 8 that performance of the linac is restored to the nominal after applying compensation scheme.

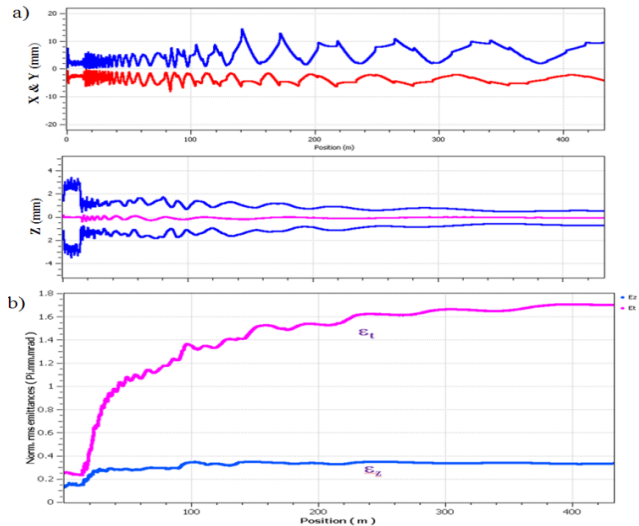

Figure 7: a) Transverse (x,y) and longitudinal r.m.s. beam envelopes and (b) emittances along SC linac in the presence of failure of the first solenoid in SSR0 section.

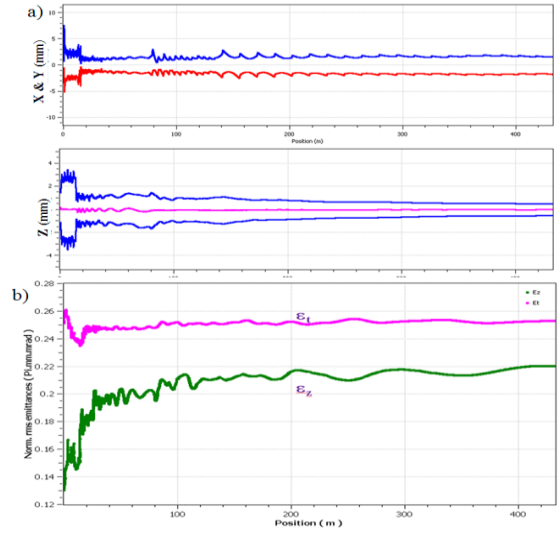

Figure 8: a) Transverse (x,y) and longitudinal(z) r.m.s. beam envelopes and (b) emittances along the SC linac in presence of failed solenoid after compensation.

## CONCLUSION

The method of compensation of failure of elements in beam line is demonstrated for most critical cases i.e. failure of first solenoid and rf cavity in SSR0 section. Results are summarized in Table2. It can be concluded that it is possible to operate linac with failed elements after applying compensation scheme.

Table 2: Beam Energy and Emittances at End of Linac for different cases.

| Parameters | No Failure | After Failure | |
|---|---|---|---|
| | | Before Compensation | After Compensation |
| Failure of first cavity in SSR0 section | | | |
| $\varepsilon_z$ (pi.mm.mrad) | 0.18 | 0.74 | 0.21 |
| $\varepsilon_t$ (pi.mm.mrad) | 0.258 | 0.27 | 0.255 |
| Eenergy (MeV) | 3028 | 3027 | 3028 |
| Failure of first solenoid in SSR0 section | | | |
| $\varepsilon_z$ (pi.mm.mrad) | 0.18 | 0.34 | 0.208 |
| $\varepsilon_t$ (pi.mm.mrad) | 0.258 | 1.7 | 0.262 |
| Eenergy (MeV) | 3028 | 3028 | 3028 |

## REFERENCES

[1] S.D. Holmes "Project X: A Multi-MW Proton Source at Fermilab", IPAC 2010.
[2] N. Solyak, et al, this Conference, MOP145.
[3] R.Duperrier et al. "CEA Saclay codes review for high intensity linac", ICCS conference, Amesterdam